\RequirePackage{lineno}
\documentclass[apj]{emulateapj}
\slugcomment{{\sc Accepted to ApJ:} June 19, 2015} 
\usepackage{epsfig}
\usepackage{epstopdf}

\usepackage{graphicx}
\usepackage{enumerate}
\usepackage[english]{babel}
\usepackage[T1]{fontenc}
\usepackage{hyperref}
\usepackage{breakurl}
\usepackage{longtable}
\usepackage{amsmath}
\usepackage{textcomp} 
\usepackage{apjfonts}
\usepackage{natbib}
\newcommand*\patchAmsMathEnvironmentForLineno[1]{%
  \expandafter\let\csname old#1\expandafter\endcsname\csname #1\endcsname
  \expandafter\let\csname oldend#1\expandafter\endcsname\csname end#1\endcsname
  \renewenvironment{#1}%
     {\linenomath\csname old#1\endcsname}%
     {\csname oldend#1\endcsname\endlinenomath}}%
\newcommand*\patchBothAmsMathEnvironmentsForLineno[1]{%
  \patchAmsMathEnvironmentForLineno{#1}%
  \patchAmsMathEnvironmentForLineno{#1*}}%
\AtBeginDocument{%
\patchBothAmsMathEnvironmentsForLineno{equation}%
\patchBothAmsMathEnvironmentsForLineno{align}%
\patchBothAmsMathEnvironmentsForLineno{flalign}%
\patchBothAmsMathEnvironmentsForLineno{alignat}%
\patchBothAmsMathEnvironmentsForLineno{gather}%
\patchBothAmsMathEnvironmentsForLineno{multline}%
}

\newcommand{\matteoemail}{\textit{matteo.cerruti@cfa.harvard.edu}}
\newcommand{\seanemail}{\textit{griffins@physics.mcgill.ca}}

\newcommand{\veritas}{\textit{VERITAS}}
\newcommand{\fermilat}{\textit{Fermi}-LAT}
\newcommand{\fermi}{\textit{Fermi}}
\newcommand{\swiftuvot}{\textit{Swift}-UVOT}
\newcommand{\uvot}{\textit{UVOT}}
\newcommand{\swiftxrt}{\textit{Swift}-XRT}
\newcommand{\xrt}{\textit{XRT}}
\newcommand{\swift}{\textit{Swift}}
\newcommand{\magic}{\textit{MAGIC}}
\newcommand{\flwo}{\textit{FLWO 48"}}
\newcommand{\hess}{\textit{H.E.S.S.}}

\newcommand{\onees}{\textit{1ES~1727+502}}

\begin{document}

\title{\veritas\  detection of $\gamma$-ray flaring activity from the BL~Lac~object \onees\ \\during bright moonlight observations}
\shorttitle{\veritas\ detection of $\gamma$-ray flaring activity from the BL~Lac~object \onees\ during bright moonlight observations}

\author{
S.~Archambault\altaffilmark{1},
A.~Archer\altaffilmark{2},
M.~Beilicke\altaffilmark{2},
W.~Benbow\altaffilmark{3},
R.~Bird\altaffilmark{4},
J.~Biteau\altaffilmark{5},
A.~Bouvier\altaffilmark{5},
V.~Bugaev\altaffilmark{2},
J.~V~Cardenzana\altaffilmark{6},
M.~Cerruti\altaffilmark{3},
X.~Chen\altaffilmark{7,8},
L.~Ciupik\altaffilmark{9},
M.~P.~Connolly\altaffilmark{10},
W.~Cui\altaffilmark{11},
H.~J.~Dickinson\altaffilmark{6},
J.~Dumm\altaffilmark{12},
J.~D.~Eisch\altaffilmark{6},
M.~Errando\altaffilmark{13},
A.~Falcone\altaffilmark{14},
Q.~Feng\altaffilmark{11},
J.~P.~Finley\altaffilmark{11},
H.~Fleischhack\altaffilmark{8},
P.~Fortin\altaffilmark{3},
L.~Fortson\altaffilmark{12},
A.~Furniss\altaffilmark{5},
G.~H.~Gillanders\altaffilmark{10},
S.~Griffin\altaffilmark{1},
S.~T.~Griffiths\altaffilmark{15},
J.~Grube\altaffilmark{9},
G.~Gyuk\altaffilmark{9},
N.~H{\aa}kansson\altaffilmark{7},
D.~Hanna\altaffilmark{1},
J.~Holder\altaffilmark{16},
T.~B.~Humensky\altaffilmark{17},
C.~A.~Johnson\altaffilmark{5},
P.~Kaaret\altaffilmark{15},
P.~Kar\altaffilmark{18},
M.~Kertzman\altaffilmark{19},
Y.~Khassen\altaffilmark{4},
D.~Kieda\altaffilmark{18},
M.~Krause\altaffilmark{8},
F.~Krennrich\altaffilmark{6},
S.~Kumar\altaffilmark{16},
M.~J.~Lang\altaffilmark{10},
G.~Maier\altaffilmark{8},
S.~McArthur\altaffilmark{20},
A.~McCann\altaffilmark{21},
K.~Meagher\altaffilmark{22},
J.~Millis\altaffilmark{23},
P.~Moriarty\altaffilmark{10},
R.~Mukherjee\altaffilmark{13},
D.~Nieto\altaffilmark{17},
A.~O'Faol\'{a}in de Bhr\'{o}ithe\altaffilmark{8},
R.~A.~Ong\altaffilmark{24},
A.~N.~Otte\altaffilmark{22},
N.~Park\altaffilmark{20},
M.~Pohl\altaffilmark{7,8},
A.~Popkow\altaffilmark{24},
H.~Prokoph\altaffilmark{8},
E.~Pueschel\altaffilmark{4},
J.~Quinn\altaffilmark{4},
K.~Ragan\altaffilmark{1},
L.~C.~Reyes\altaffilmark{25},
P.~T.~Reynolds\altaffilmark{26},
G.~T.~Richards\altaffilmark{22},
E.~Roache\altaffilmark{3},
M.~Santander\altaffilmark{13},
G.~H.~Sembroski\altaffilmark{11},
K.~Shahinyan\altaffilmark{12},
A.~W.~Smith\altaffilmark{18},
D.~Staszak\altaffilmark{1},
I.~Telezhinsky\altaffilmark{7,8},
J.~V.~Tucci\altaffilmark{11},
J.~Tyler\altaffilmark{1},
A.~Varlotta\altaffilmark{11},
S.~Vincent\altaffilmark{8},
S.~P.~Wakely\altaffilmark{20},
A.~Weinstein\altaffilmark{6},
R.~Welsing\altaffilmark{8},
A.~Wilhelm\altaffilmark{7,8},
D.~A.~Williams\altaffilmark{5},
B.~Zitzer\altaffilmark{27} (the VERITAS Collaboration)
and Z.~D.~Hughes\altaffilmark{5}
}

\altaffiltext{1}{Physics Department, McGill University, Montreal, QC H3A 2T8, Canada}
\altaffiltext{2}{Department of Physics, Washington University, St. Louis, MO 63130, USA}
\altaffiltext{3}{Fred Lawrence Whipple Observatory, Harvard-Smithsonian Center for Astrophysics, Amado, AZ 85645, USA}
\altaffiltext{4}{School of Physics, University College Dublin, Belfield, Dublin 4, Ireland}
\altaffiltext{5}{Santa Cruz Institute for Particle Physics and Department of Physics, University of California, Santa Cruz, CA 95064, USA}
\altaffiltext{6}{Department of Physics and Astronomy, Iowa State University, Ames, IA 50011, USA}
\altaffiltext{7}{Institute of Physics and Astronomy, University of Potsdam, 14476 Potsdam-Golm, Germany}
\altaffiltext{8}{DESY, Platanenallee 6, 15738 Zeuthen, Germany}
\altaffiltext{9}{Astronomy Department, Adler Planetarium and Astronomy Museum, Chicago, IL 60605, USA}
\altaffiltext{10}{School of Physics, National University of Ireland Galway, University Road, Galway, Ireland}
\altaffiltext{11}{Department of Physics and Astronomy, Purdue University, West Lafayette, IN 47907, USA}
\altaffiltext{12}{School of Physics and Astronomy, University of Minnesota, Minneapolis, MN 55455, USA}
\altaffiltext{13}{Department of Physics and Astronomy, Barnard College, Columbia University, NY 10027, USA}
\altaffiltext{14}{Department of Astronomy and Astrophysics, 525 Davey Lab, Pennsylvania State University, University Park, PA 16802, USA}
\altaffiltext{15}{Department of Physics and Astronomy, University of Iowa, Van Allen Hall, Iowa City, IA 52242, USA}
\altaffiltext{16}{Department of Physics and Astronomy and the Bartol Research Institute, University of Delaware, Newark, DE 19716, USA}
\altaffiltext{17}{Physics Department, Columbia University, New York, NY 10027, USA}
\altaffiltext{18}{Department of Physics and Astronomy, University of Utah, Salt Lake City, UT 84112, USA}
\altaffiltext{19}{Department of Physics and Astronomy, DePauw University, Greencastle, IN 46135-0037, USA}
\altaffiltext{20}{Enrico Fermi Institute, University of Chicago, Chicago, IL 60637, USA}
\altaffiltext{21}{Kavli Institute for Cosmological Physics, University of Chicago, Chicago, IL 60637, USA}
\altaffiltext{22}{School of Physics and Center for Relativistic Astrophysics, Georgia Institute of Technology, 837 State Street NW, Atlanta, GA 30332-0430}
\altaffiltext{23}{Department of Physical Sciences and Engineering, Anderson University, 1100 East 5th Street, Anderson, IN 46012}
\altaffiltext{24}{Department of Physics and Astronomy, University of California, Los Angeles, CA 90095, USA}
\altaffiltext{25}{Physics Department, California Polytechnic State University, San Luis Obispo, CA 94307, USA}
\altaffiltext{26}{Department of Applied Science, Cork Institute of Technology, Bishopstown, Cork, Ireland}
\altaffiltext{27}{Argonne National Laboratory, 9700 S. Cass Avenue, Argonne, IL 60439, USA\\}

\begin{abstract}
During moonlit nights, observations with ground-based Cherenkov telescopes at very high energies (VHE, $E>100$ GeV) are constrained since the photomultiplier tubes (PMTs) in the telescope camera are extremely sensitive to the background moonlight. Observations with the \veritas\ telescopes in the standard configuration are performed only with a moon illumination less than 35$\%$ of full moon. Since 2012, the \veritas\ collaboration has implemented a new observing mode under bright moonlight, by either reducing the voltage applied to the PMTs (reduced-high-voltage configuration, RHV), or by utilizing UV-transparent filters. While these operating modes result in lower sensitivity and increased energy thresholds, the extension of the available observing time is useful for monitoring variable sources such as blazars and sources requiring spectral measurements at the highest energies. In this paper we report the detection of $\gamma$-ray flaring activity from the BL Lac object \onees\ during RHV observations. This  detection represents the first evidence of VHE variability from this blazar. The integral flux is $(1.1\pm0.2)\times10^{-11}\mathrm{cm^{-2}s^{-1}}$ above 250 GeV, which is about five times higher than the low-flux state. The detection triggered additional \veritas\ observations during standard dark-time. Multiwavelength observations with the \flwo\ telescope, and the \swift\ and \fermi\ satellites are presented and used to produce the first spectral energy distribution (SED) of this object during $\gamma$-ray flaring activity. The SED is then fitted with a standard synchrotron-self-Compton model, placing constraints on the properties of the emitting region and of the acceleration mechanism at the origin of the relativistic particle population in the jet. \\  
\end{abstract}


\section{Introduction}
\label{section1}

{\let\thefootnote\relax\footnotetext{Send off-print requests to:\\ Matteo Cerruti (\matteoemail)\\ and Sean Griffin (\seanemail)\vspace{0.3cm}}}
\setcounter{footnote}{0}

Blazars are a class of radio-loud active galactic nuclei (AGN) characterized by a broadband nonthermal continuum from radio to $\gamma$-rays, extreme variability and a high degree of polarization. In the framework of the unified AGN model \citep[see e.g.][]{Urry95} they are considered AGN whose relativistic jet is aligned with the line of sight. The blazar spectral energy distribution (SED) is thus dominated by the emission from the jet, enhanced by relativistic effects. In the $\gamma$-ray sky, blazars are the dominant AGN class, representing $97\%$ of the \fermilat\ AGN \citep[between 100 MeV and 100 GeV,][]{2LAC}, and $>90\%$ of extragalactic sources detected at very high energies (VHE; $E > 100$ GeV) by ground-based imaging atmospheric-Cherenkov telescopes (IACTs)\footnote{see \url{http://tevcat.uchicago.edu} for a regularly updated list of known TeV sources.}.\\

The blazar class is composed of two subclasses, flat-spectrum radio quasars (FSRQ) and BL Lacertae objects, depending on the presence (in the former) or absence (in the latter) of emission lines in their optical/UV spectrum \citep[the threshold between the two subclasses is an emission line equivalent width of 0.5 nm, see e.g.][]{Angel80}. The two subclasses are also characterized by different luminosity and redshift distributions. The FSRQs are on average brighter and more distant than BL Lac objects \citep[see e.g.][]{Padovani92}. This dichotomy in the blazar class reflects the dichotomy observed in the radio-galaxy population: FSRQs are considered the blazar version of the Fanaroff-Riley II \citep[FR II,][]{FR74} radio-galaxies, while BL Lac objects are believed to represent the blazar version of FR I.  The two subclasses share the same SED shape: a broadband continuum from radio to gamma-rays, composed of two separate bumps, peaking in IR-to-X-rays and MeV-to-TeV, respectively. While FSRQs are generally characterized by a lower frequency of the first peak, BL Lac objects show a variety of peak frequencies, and are further classified \citep[see e.g.][]{Padovani95} into low-frequency-peaked BL Lacs (LBL, with $\nu_{peak}$ in infrared) and high-frequency-peaked BL Lacs (HBL, with $\nu_{peak}$ in UV and beyond). When $\nu_{peak}$ is located in the optical/UV the object is often classified as an intermediate-frequency-peaked BL Lac (IBL).\\

The position of the second peak is related to the position of the first one, as shown by observations in the MeV-GeV energy band with \fermilat. While FSRQs and LBLs present a peak at MeV/GeV, HBLs show a $\gamma$-ray component peaking at higher energies, often above the \fermilat\ energy band \citep{Abdo10}. \citet{Fossati98} proposed the existence of an anti-correlation between the blazar luminosity and $\nu_{peak}$ (the so-called blazar sequence), although there is not a general consensus on this point. More recently \citet{Meyer11} extended this sequence into a more general "blazar envelope". 
The subclass of HBLs, even though the least luminous among the other blazar subclasses, is the brightest one at VHE, and the majority of VHE blazars are indeed HBLs \citep[see e.g.][for a recent review]{Gunes13}.\\ 

The study of VHE blazars is complicated by two observational characteristics of blazars themselves: their broadband emission and their rapid variability. A VHE detection on its own, sampling only a small part of the nonthermal continuum, does not provide sufficient information about the underlying blazar physics. Rather, strictly simultaneous multiwavelength (MWL) campaigns are required to constrain the blazar SED.
The launch of the \fermi\ satellite and its monitoring capabilities have had a large impact on blazar physics, providing for the first time long-term light curves of hundreds of blazars in $\gamma$-rays, and assuring simultaneous observations of VHE blazars in the MeV/GeV energy band.  \\

Observations by IACTs are limited by the high sensitivity of the camera photomultiplier tubes (PMTs), which in their standard configuration experience higher noise, higher current level, and accelerated aging when operated in bright moonlight conditions (moon illumination $\geq 35\%$ of full moon). This constraint particularly affects blazar studies, for example, limiting the organization of MWL campaigns, or prohibiting follow-up observations of a flare.  Several experiments have successfully performed observations of Cherenkov light under bright moonlight conditions \citep{Weekes86, Whipple95, Whipple97, Artemis01, Hegra01, Hegra03}. Of the current generation of IACTs, only \magic\  and \veritas\ \citep[see][]{Magicmoon00, Rico07, Magicmoon} observe under moderate moonlight ($< 35\%$ moon illumination) which significantly improves the duty cycle of the observatories. \textit{FACT}, whose camera is composed of solid-state Geiger-mode avalanche photodiodes (also called silicon-photomultipliers), instead of PMTs \citep{FACT2013}, is also capable of observing under moonlight.\\

In 2012, the \veritas\ collaboration began a new program of observing under bright moonlight, by applying reduced high voltage (RHV) to lower the PMT gain, or by utilizing UV-transparent filters. The Schott Glass UG-11 filters' bandpass is from $275$ nm to $375$ nm, reducing moonlight by a factor of ten and Cherenkov radiation by a factor of three. The details of the observing strategy have been presented at several conferences \citep{Dumm13, Staszak14} and will be discussed in an upcoming \veritas\ publication (2015, in preparation). In this paper, we concentrate on the capabilities of the \veritas\ telescope array in the RHV configuration. We also present the detection, during May 2013, of VHE emission from the blazar \onees\ at a flux of roughly five times the archival VHE flux measured by \magic\ \citep{MAGICpaper}. The high-flux state was initially detected during bright moonlight observations, which represents an innovation for \veritas.\\

 The blazar \onees\ \citep[from the \textit{Einstein} Slew Survey Catalog,][]{1ES}, also known as \textit{I~Zw~187} or \textit{OT~546}, is a nearby \citep[$z = 0.055$,][]{Oke78} HBL, discovered as a $\gamma$-ray source by \fermi\ \citep{1FGL}, and as a VHE source by \magic\ \citep{MAGICpaper}. Radio observations show a compact core-jet morphology \citep{Laurent93, Koll96, Pushkarev12}, typical of blazars, with an apparent jet opening angle of 11$^\circ$ \citep{Linford12}. Past optical monitoring shows only moderate variability \citep{Pica88, Fiorucci96} and the presence of a weak 300~nm bump \citep{Bregman82}, quite unusual in BL Lac objects. 
 Prior to being detected by \magic\ as a VHE source, \onees\ was observed by both \textit{HEGRA} \citep{HegraUL} and the \textit{WHIPPLE} 10-m telescope \citep{WhippleUL}, with no detection.\footnote{A flux upper limit of $8.6\%$ ($9\%$) Crab Nebula units above $300$ ($940$) GeV was measured by \textit{Whipple} (\textit{Hegra}).} The \magic\ collaboration reported an integral flux of $(5.5 \pm 1.4)\times 10^{-12}$ cm$^{-2}$ s$^{-1}$ above $150$ GeV (corresponding to  
 $(2.1 \pm 0.4)\%$ of the Crab Nebula flux\footnote{Based on the \magic\ measurement of the Crab Nebula \citep{MAGICCrab} }) and a spectral index of $2.7 \pm 0.5$ \citep[see][]{MAGICpaper}.\\

This paper is organized as follows: in Section \ref{section2} we describe the capabilities of the \veritas\ instrument in the RHV configuration; in Section \ref{section3} we describe the \veritas\ observations of \onees, in both RHV and standard configurations; the details of the MWL campaign triggered by the detection of the high-flux state are reported in Section \ref{section4}; in Section \ref{section5} we model the  SED using a standard one-zone synchrotron-self-Compton model, and in Section \ref{section6} we discuss the results obtained from the MWL campaign.\\

\section{\veritas\ instrument} 
\label{section2}

\veritas\ is an array of four imaging atmospheric Cherenkov telescopes located at the Whipple Observatory in southern Arizona (31\textdegree40'N, 110\textdegree57'W) at an altitude of 1.3~km above sea level. Each telescope is of Davies-Cotton design \citep{1957SoEn....1...16D} with a 12-m diameter reflector, and the array is arranged in a diamond configuration with $\sim$ 100~m to a side. Each reflector comprises 345 identical hexagonal mirror facets and has a collection area of 110~m$^2$.\\

Each \veritas\ telescope is instrumented with a camera made up of 499 PMTs each with a 0.15\textdegree\ field of view (FoV), for a total FoV diameter of 3.5\textdegree. The PMT signals are digitized by 500~megasamples-per-second flash analog-to-digital converters (FADCs). \veritas\ employs a 3-level trigger system \citep{2008ICRC....3.1539W, 2013arXiv1307.8360Z} and has an array trigger rate of $\sim$~450~Hz.\\

\veritas\ has an energy resolution of 15\%, and a single-event angular resolution of 0.1\textdegree\ at 1~TeV. During standard observations (i.e. when the moon is $< 35\%$ illuminated) a source with an integrated flux of 1\% of the Crab Nebula flux can be detected at the $5\sigma$ level in $\sim$~25~hours, and a 5\% Crab source in less than 2 hours. More information on the \veritas\ array can be found in \cite{Holder06, 2011ICRC...12..137H} and \cite{2013arXiv1308.4849D}.\\

In the RHV observation mode, the PMT voltages are reduced to 81\% of their standard values during dark-sky observations, allowing \veritas\ to operate when the moon is 35-65\% illuminated. 
This reduces the absolute gain of each PMT by a factor of $\sim 3.2$. Observing in RHV mode increases the yearly exposure of \veritas\ by $\sim13\%$ above 250~GeV. 
For RHV data, systematic errors induced by uncertainty in the atmosphere, telescope optical point spread function, and mirror reflectivity are the same as those for standard data. 
In terms of sensitivity, a standard analysis of \veritas\ observations of the Crab Nebula taken under dark skies at small zenith angles yields a sensitivity of $\sim 35\sigma/\sqrt{\mathrm{hr}}$ and an analysis energy threshold of $\sim $~170~GeV.
In comparison, Crab data taken in RHV mode under moonlight also have a sensitivity $\sim 35\sigma/\sqrt{\mathrm{hr}}$ albeit with a higher energy threshold ($\sim$~200~GeV). These values assume a Crab-like spectral index. Note that \veritas\ event-selection cuts are optimized for sensitivity and not energy threshold. The cuts used in this RHV analysis would allow for a 5\% Crab source to be detected in less than 2 hours.\\

\section{\veritas\ observations of \onees\ }
\label{section3}

 The \veritas\ observations that allowed for the detection of \onees\ were taken between May 1, 2013 (MJD 56413) and May 7, 2013 (MJD 56419). 
 There were additional data taken on May 18, 2013 (MJD 56430) that did not result in a detection. 
 After quality selection, approximately 6 hours of data remain. Of these, 3 hours were taken in RHV mode on the first two nights of the exposure. 
 All observations were made in ``wobble'' mode, wherein the telescopes were pointed 0.5\textdegree\ away from the target to allow simultaneous measurements of the target and background regions \citep{1994APh.....2..137F}, and all data in the 2013 dataset were taken at elevations between 65$^\circ$ and 70$^\circ$. \\

Two sets of gamma-hadron separation cuts, each optimized \emph{a priori} 
on data taken on the Crab Nebula where the gamma-ray excess has been 
scaled down to 5\%, were used in the analysis of the 2013 dataset. 
For the standard-voltage subset, only events with images in three 
or more telescopes were used, whereas in the RHV data subset 
two-telescope events were used and a smaller cut was made on the brightness 
of the shower images to account for the fact that the camera gain was 
reduced. The motivation for using these two sets of cuts is to approximately match the energy threshold of the analyses, resulting in a threshold of $\sim$~220~GeV for both the RHV and standard voltage datasets.\\

The analysis of the complete 2013 dataset used the ``reflected-region'' 
background model \citep{2001A&A...370..112A} resulting in 159 ON events 
and 850 OFF events with a background normalization factor $\alpha$ of 0.077, 
yielding a detection significance of $9.3\sigma$ \citep[][Eq. 17]{1983ApJ...272..317L} 
and an average gamma-ray-like event rate of $0.25\pm0.04$ counts/minute 
with a background event rate of $0.18\pm0.02$ counts/minute.\\

Prior to the detection of \onees, \veritas\ observed the object for 8.6 hours between 2007 and 2009. There was no detection in these data; the upper limit (99\% confidence level) is $2.6\times 10^{-12}~\mathrm{cm^{-2} s^{-1}}$ ($2.5\%$ Crab) above 350~GeV assuming the spectral index from \citet{MAGICpaper}. This is consistent with the blazar's low-flux state as detected by \magic\  in \cite{MAGICpaper}.\\

\begin{table*}[htbp]
    \caption{The data points for the \veritas\ observed spectrum of \onees.}
    \label{tbl:spectralPoints}
    \begin{center}
    
    \begin{tabular}{ccccccc}
    \hline\hline
    $<\mathrm{E}>$   & $\mathrm{E_{low}}$    & $\mathrm{E_{high}}$   & Flux or UL (95\% CL) & Flux Error  & Excess\tablenotemark{a} & Sig.      \\
    $[$TeV$]$ & $[$TeV$]$ & $[$TeV$]$ & $[\mathrm{cm^{-2} s^{-1} TeV^{-1}}]$ & $[\mathrm{cm^{-2} s^{-1} TeV^{-1}}]$ &  & $[\sigma]$ \\\hline
    0.32         & 0.25         & 0.40         & $3.3\times 10^{-11}$       & $9.6\times 10^{-12}$       & 25.8   & 4.1               \\
    0.50         & 0.40         & 0.63         & $1.1\times 10^{-11}$       & $3.1\times 10^{-12}$       & 19.2   & 4.4               \\
    0.79         & 0.63         & 1.00         & $5.9\times 10^{-12}$       & $1.4\times 10^{-12}$       & 21.0   & 7.0               \\
    1.26         & 1.00         & 1.59         & $1.4\times 10^{-12}$       & $5.0\times 10^{-13}$       & 8.8   & 4.5               \\
    2.00         & 1.59         & 2.51         & $4.1\times 10^{-13}$       & n/a                        & 0.8    &                  \\
    3.16         & 2.51         & 3.98         & $3.4\times 10^{-13}$       & n/a                        & 2.0$^*$&                  \\
    5.01         & 3.98         & 6.31         & $1.6\times 10^{-13}$       & n/a                        & 1.0$^*$    &                  \\
    \hline\hline
    \end{tabular}
    \tablenotetext{1}{The $^*$ indicates that the bin contained no OFF events. \vspace{0.5cm}}
    \end{center}
\end{table*}

\begin{figure*}[hbt]
\begin{center}

\includegraphics[width=420pt]{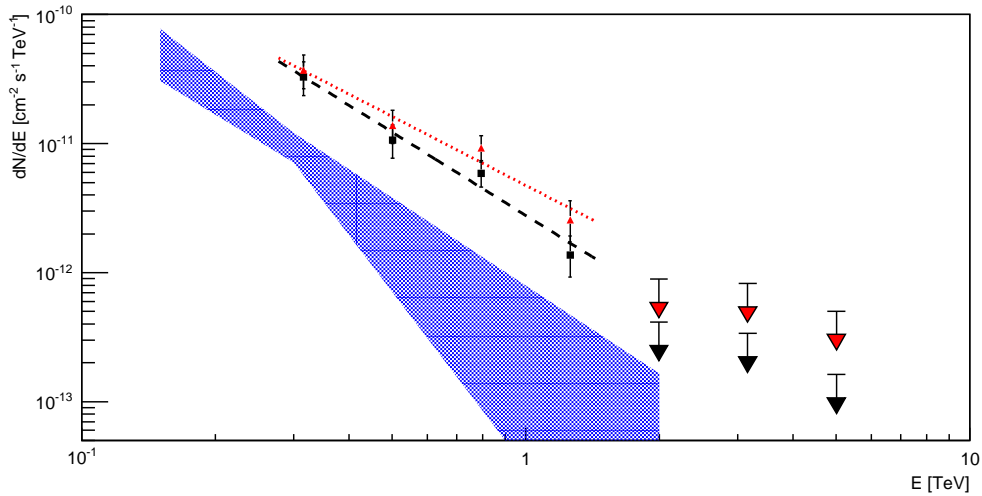}
\caption{Spectrum for the complete \onees\ dataset. Power-law fits to both the uncorrected (black, squares) and EBL-corrected (red, triangles) points are provided. The fitted values are given in the text. The blue shaded region represents the result from \magic\ \citep{MAGICpaper}. Note that the first upper limit is below the extrapolation of the fit. This may be indicative of a cutoff, however, the number of events in this bin is low (1 ON and 2 OFF events with $\alpha=0.08$ at 2~TeV), thus, it is also possible that this is simply a downward statistical fluctuation.}
\label{fig:veritasSpectrum}
\end{center}
\end{figure*}

The fitted position of the excess events is $(\alpha,\delta)_\mathrm{J2000} 
= ( 17^h~28^m~4.1^s \pm 7.5^s_\mathrm{stat} \pm 5.2^s_\mathrm{sys}, 
+50^\circ~13'~60'' \pm 1'~10''_\mathrm{stat} \pm 50''_\mathrm{sys} )$ 
and is within $0.04^\circ$ ($2.4'$) of the catalog position 
$(\alpha,\delta) = (17^h~28^m~18.624^s, +50^\circ~13'~10.416'')$
 given in \cite{1998AJ....116..516M}. The \veritas\ catalog name 
 for this source is VER~J1728+502. The excess seen from the target
  is consistent with a point source.\\

The reconstructed VHE spectrum is shown in Fig.~\ref{fig:veritasSpectrum}. 
In the same figure we show data points corrected for absorption by extragalactic background light (EBL) using the model in \cite{Franceschini08} (for z=0.055). 
Both are well fitted with a power-law function between 0.25 TeV and 1.6~TeV; the $\chi^2 / NDF$  for the observed and EBL-corrected spectra are $1.66/2$ ($P \sim$  44\%) and $1.69/2$ ($P \sim$ 43\%), respectively. 
The observed spectrum is given by
\begin{equation*}
  \frac{dN}{dE} = (7.8 \pm 1.1 )\times 10^{-12}  \left(\frac{E}{620~\mathrm{GeV}}\right)^{-2.1 \pm 0.3} \mathrm{cm^{-2}s^{-1}TeV^{-1}},
\end{equation*}
and the EBL-corrected (\emph{intrinsic}) spectrum is given by
\begin{equation*}
  \frac{dN}{dE} = (1.1 \pm 0.2)\times 10^{-11}  \left(\frac{E}{620~\mathrm{GeV}}\right)^{-1.8 \pm 0.3} \mathrm{cm^{-2}s^{-1}TeV^{-1}}.
\end{equation*}
All errors in the aforementioned fits are statistical. 
The cumulative systematic errors on the flux normalizations and spectral indices are conservatively estimated to be 30\% and $\pm0.3$, respectively. This is 50\% larger than the standard \veritas\ systematic uncertainties. The increase does not have a significant impact on the following discussion; a detailed discussion of systematic uncertainties during moonlight observations will be discussed in the VERITAS moonlight paper (2015, in preparation).
The numerical values for the spectral points are given in Table~\ref{tbl:spectralPoints}. \\

The light curve for the \veritas\ observations is shown in the top panel of Fig.~\ref{1727LC}. 
The peak integrated flux above 250~GeV is $(1.6 \pm 0.4)\times10^{-11}\mathrm{cm^{-2}s^{-1}}$, which corresponds to 9.5\% of the Crab Nebula flux.\footnote{Based on the power law in \cite{1998ApJ...503..744H} extrapolated to 250~GeV.} 
The upper limit is at the 95\% confidence level and represents $9.4\times10^{-12}\mathrm{cm^{-2}s^{-1}}$ (5.6\%~Crab). 
The light curve between May~01 and May~07 (MJD 56413-56419) can be fitted with a constant resulting in a flux of $(1.1\pm0.2)\times10^{-11}\mathrm{cm^{-2}s^{-1}}$ (6.3\%~Crab) and a $\chi^2/NDF = 4.73/3$ ($P \sim$ 19\%). \\

The integral flux above 250 GeV corresponds to about five times the flux measured by \magic, and represents the first evidence of VHE variability in \onees. The \veritas\ observations between May~01 and May~07 are consistent with a constant flux. However, when including the measurement on May~18 in the fit, the resulting $\chi^2/NDF$ is $12.0/4$ ($P \sim$ 1.8\%). Thus, a constant flux is excluded at the 2.4$\sigma$ level, indicating that the flare may have ended at some point after the last \veritas\ detection on May~07.\\

\section{Multiwavelength observations of \onees\ }
\label{section4}

\subsection{\fermilat\ }
\label{fermisection}
The blazar \onees\ was first detected as a $\gamma$-ray source by \fermilat\ \citep{FermiTeV}, and included in both the first and the second \fermi\ catalogs \citep{1FGL,2FGL}. In the latter, it is named \textit{2FGL J1728.2+5015} and is detected with a significance of 9.0$\sigma$. Its spectrum is parametrized by a power-law function with index $\Gamma_{2FGL}=1.83 \pm 0.13$ and differential flux $\Phi_{2FGL}=(9.5 \pm 1.6)\times10^{-14}$ cm$^{-2}$ s$^{-1}$ MeV$^{-1}$, estimated at the decorrelation energy $E_{0;2FGL}=2935$ MeV. The source has also been included in the catalog of hard \fermilat\ sources \citep[1FHL,][]{1FHL}, with the name 1FHL J1728.3+5014 and an estimated power-law index between 10 GeV and 500 GeV of $\Gamma_{1FHL}=1.67 \pm 0.34$, consistent with the 2FGL result. There is no evidence of curvature in its \fermilat\ spectrum above 10 GeV.\\

A new analysis of the \fermilat\ Pass7 \citep[see][]{PassSeven} data has been performed using version \textit{v9r32p5} of the \textit{ScienceTools}\footnote{See \url{http://fermi.gsfc.nasa.gov/ssc/data/analysis/software/}}. Only photons passing the {\small SOURCE} class filter and located within a square region of side length $40^\circ$ centered on \onees\ were selected. We used photons with energies from 100 MeV to 300 GeV and included observations from August 4, 2008 (the beginning of the \fermi\ mission) to August 1, 2013. Following the prescriptions of the \fermilat\ Collaboration,\footnote{See \url{http://fermi.gsfc.nasa.gov/ssc/data/analysis/LAT_caveats.html}} the analysis was performed using the binned likelihood method, and the \textit{P7REP$\_$SOURCE$\_$V15} instrumental response function. Data were filtered considering only zenith angles lower than 100$^\circ$ and rocking angles lower than 52$^\circ$. The Galactic diffuse component and the isotropic background were modeled using the templates provided by the \fermi\ team, namely \textit{gll$\_$iem$\_$v05.fits} and \textit{iso$\_$source$\_$v05.txt}. The likelihood analysis was performed using \textit{gtlike} and all the 2FGL sources present in the region of interest were included in the model, using the script \textit{make2FGLxml.py}.\footnote{See \url{http://fermi.gsfc.nasa.gov/ssc/data/analysis/user/}} For every source the spectral function (log parabola or power law) used in the 2FGL catalog was adopted, with parameters free to vary. The model also included sources outside the region of interest, up to a distance of 35$^\circ$, with spectral parameters frozen to the 2FGL values. The normalizations of the models for the Galactic diffuse component and the isotropic background were left free to vary during the fit.\\

Using this dataset, \onees\ was detected with a test-statistic \citep[as defined in][]{1FGL} of $219$, corresponding to a significance of $\sim$ 14$\sigma$. Its emission is parametrized by a power-law function with index $\Gamma_{0.1-300 GeV}=1.91 \pm 0.08$ and differential flux $\Phi_{0.1-300 GeV}=(2.1 \pm 0.2)\times10^{-13}$ cm$^{-2}$ s$^{-1}$ MeV$^{-1}$, estimated at the decorrelation energy $E_{0;0.1-300 GeV}=2136$ MeV, consistent with the 2FGL result. The presence of curvature in the spectrum was tested by replacing the power-law model with a log-parabolic one: the value of the curvature parameter $\beta$ is consistent with zero, supporting the power-law assumption. In the residual map there is no evidence of additional sources besides the ones included in the 2FGL catalog. Spectral points were computed for five different spectral bins, using the \textit{python} scripts prepared by the \fermilat\  Collaboration,\footnote{See \url{http://fermi.gsfc.nasa.gov/ssc/data/analysis/scitools/python_tutorial.html}} assuming a fixed spectral index $\Gamma_{0.1-300 GeV}=1.91$. If the significance is lower than 5$\sigma$, 95$\%$ upper limits are provided. The bow-tie of the best-fit \fermi\ measurement, as well as the spectral data, are plotted in Fig. \ref{figSED}.\\

The 2FGL catalog reports a variability index of 18 \citep[see][for details]{2FGL}, indicating a lack of variability. A test for variability in the MeV-GeV band has been performed using aperture photometry: only photons detected in a 1$^\circ$-radius region around the blazar have been considered, and count-rate light curves have been produced using \textit{gtbin} with time bins of  one day, one week, one month and three months. All four light curves are consistent with a constant flux.\\

A second, unbinned, analysis was performed in order to produce a measurement in the GeV energy band spanning the same interval as \veritas\ observations. Only \fermilat\ observations taken between May 1 and May 7, 2013 (MJD 56413 and 56419), inclusive, are considered. In this case the region of interest is a circle of 20$^\circ$ radius centered on \onees. The model includes all the sources used in the binned analysis of five years of \fermilat\ data, with spectral parameters frozen if the source is outside the region of interest. \onees\ is not detected by \fermilat\ in this very short period and only upper limits on its GeV emission can be computed. They are estimated at the 95\% confidence level for three different energy bands (0.1-1; 1-10 and 10-300 GeV), and are plotted in Fig. \ref{figSED}.\\

\begin{figure}[hbtp]
\begin{center}
\includegraphics[width=260pt]{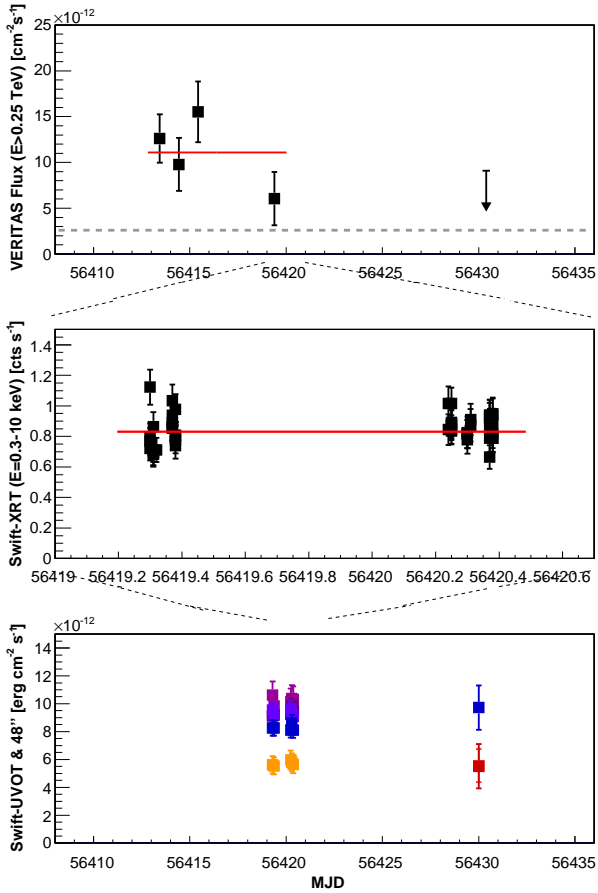}
\vspace{0.1cm}
\caption{Light curve of \onees\ during the May 2013 multiwavelength campaign. The time axis is in Modified Julian Days. For reference, MJD 56415 corresponds to May 3, 2013. \textbf{Top panel:} \veritas\ flux light curve, measured above 250 GeV and in daily bins. The errors are statistical only, and the first two data points were taken in RHV mode. The red line corresponds to the average \veritas\ flux, while the dotted black line corresponds to the archival VHE flux reported by \magic. \textbf{Central panel:} \swiftxrt\ light curve, expressed in counts per second above 0.3 keV. The red line corresponds to the average X-ray flux during this campaign. Note that in this panel the X-axis is zoomed on MJDs 56419 and 56420. \textbf{Bottom panel:}  \swiftuvot\ (on MJD 56419-20) and \flwo\ (on MJD 56430) light curve, expressed as $\nu F_\nu$. Data have been corrected for both Galactic absorption and host-galaxy contamination. Red points correspond to i' and r' filters (the two are almost superposed on MJD 56430), orange to V, blue to B, violet to U, magenta to the UVW1,UVW2 and UVM2. \label{1727LC}}
\end{center}
\end{figure}

\subsection{\swiftxrt\ }
\label{swiftxrt}
The detection of the high-flux state at VHE by \veritas\ triggered X-ray and UV observations by the \swift\ satellite \citep{Swift}, which observed \onees\ on May 7 and May 8, 2013 (MJD 56419 and 56420) for a total live time of 6.8 ks. The X-Ray Telescope \citep[XRT,][]{XRT} performed observations in windowed-timing mode, where only the 200 central columns of the detector are read, improving the time resolution of the instrument \citep[see][]{Cusumano12}. The data analysis is performed using \textit{HEASoft} (version 6.13). Cleaned event files are produced using default screening criteria. Images, light curves, and spectra are extracted (using \textit{XSelect}, version 2.4b) from a box with height equal to 10 bins and length equal to 40 bins for both the source and the background region. \\

The \xrt\ light curve above $0.3$ keV (in counts per second), corrected for the exposure\footnote{See \url{http://www.swift.ac.uk/analysis/xrt/lccorr.php}} and for background, is shown in Fig. \ref{1727LC}. No significant variability is detected within a single observation, nor between the two observations: a fit of the light curve with a constant function yields a $\chi^2/NDF$ value of 714/682 (chance probability of $19\%$), consistent with the flux being steady (within the statistical uncertainties). The mean count rate is $0.83\pm0.12$ counts per second. The rate is low enough to avoid any significant pile-up effect in the detector. The data shown in Fig. \ref{1727LC} have been rebinned for plotting purposes.\\

The spectral analysis is performed using \textit{XSpec} (version 12.8.0). Given the lack of variability, and in order to improve the statistics, the data are summed (using \textit{mathpha}) and rebinned (using \textit{grppha}) assuming a minimum of 50 counts per bin. The response files provided by the \swift\ science team are used, while the ancillary response files are computed using \textit{xrtmkarf}. Data below $0.3$ keV are excluded, and the last significant spectral bin extends up to $8$ keV.\\

The first spectral model tested is a simple absorbed power law\footnote{The Galactic absorption is computed using the \textit{tbnew} model, an updated version of \textit{tbabs} \citep{tbabs}. See \url{http://pulsar.sternwarte.uni-erlangen.de/wilms/research/tbabs/}}. The neutral absorption is fixed to the Galactic value $N_H=2.75\times 10^{20}$ cm$^{-2}$, as measured by \citet{Dickey90}\footnote{See \url{http://heasarc.gsfc.nasa.gov/cgi-bin/Tools/w3nh/w3nh.pl}}. The best-fit result is $\Gamma=2.22\pm0.04$, with a normalization factor $C=(4.9\pm0.1)\times10^{-3}$ cm$^{-2}$ s$^{-1}$ keV$^{-1}$, estimated at 1 keV. However, the $\chi^2/NDF$ is 108/86, and significant residuals are seen above 5 keV. The fit is significantly improved (F-test probability equal to $8\times10^{-5}$) if a break is added to the model, considering an absorbed broken-power-law function. The best-fit parameter values are $\Gamma_1=2.01^{+0.10}_{-0.10}$, $\Gamma_2=2.44^{+0.10}_{-0.10}$, $E_{break}=1.21^{+0.24}_{-0.20}$ keV, and normalization $C=(5.3^{+0.2}_{-0.2})\times10^{-3}$ cm$^{-2}$ s$^{-1}$ keV$^{-1}$. The $\chi^2/NDF$ value is 86/84. The measured (i.e. not corrected for Galactic absorption) flux in the energy range 0.3-8 keV is $F_X=(2.06^{+0.04}_{-0.05})\times 10^{-11}$ erg cm$^{-2}$ s$^{-1}$.\\

The observed spectral break could be intrinsic or due to an underestimation of the neutral absorption: in particular, additional absorption due to the blazar host galaxy should be studied \citep[for a detailed analysis of this effect see][]{Perlman05, Amy2013}. This hypothesis is tested by fitting the data with an absorbed power-law function, letting the value of $N_H$ to vary. The amount of absorption required to mimic the break is $N_H=(6.5^{+2.0}_{-1.7})\times 10^{20}$ cm$^{-2}$, which is about 2.4 times the Galactic absorption estimated by \citet{Dickey90}. The $\chi^2/NDF$ value of this fit is however only 95/85 (chance probability of 0.21; the F-test probability between this model and the broken-power-law fit is equal to $4\times10^{-3}$), suggesting that the intrinsic origin of the break is preferred. The spectrum shown in Fig. \ref{figSED} is the one computed assuming a broken-power-law function, corrected for the Galactic absorption from \citet{Dickey90}.\\

\swiftxrt\ also observed \onees\ during 2010\footnote{See as well the automatic \swiftxrt\ preliminary analysis, which includes the two observations presented here: \url{http://www.swift.psu.edu/monitoring/}}  and those observations were analyzed by both \citet{Massaro11} and \citet{MAGICpaper}. The count rate during these previous observations is lower (around 0.6 counts per second, 73\% of the 2013 rate), suggesting that the VHE high-flux state seen by \veritas\ was associated with a higher-flux X-ray state. While \citet{MAGICpaper} performed a simple power-law fit, \citet{Massaro11} confirmed the deviation from a power law, successfully fitting the \swiftxrt\ data with a log-parabolic function. \\

\subsection{\swiftuvot\ }
\label{swiftuvot}
The \uvot\ telescope \citep{UVOT}, on board the \swift\ satellite, observed \onees\ at optical and ultraviolet wavelengths, simultaneously with \xrt. All measurements were performed using the six available filters: V and B in the optical, and U, UVW1, UVM2 and UVW2 in the ultraviolet. A circular aperture with radius 5" is used for the source, while the background is estimated from a larger region with radius 15". The flux is estimated using \textit{uvotmaghist} (version 1.1). As in the X-ray observations, no variability was detected either within or between the two \uvot\ observations.\\

The Galactic extinction is taken into account assuming $E_{B-V}=0.037$, a value consistent with the $N_H$ value used for the X-ray analysis \citep{Jenkins74}. \citet{Schlafly11} estimated a similar value of $E_{B-V}=0.026$. Assuming $E_{B-V}=0.026$ instead of $0.037$ affects the flux estimations by between 3\% and 9\%: this additional systematic uncertainty is included in the \uvot\ error bars plotted in Fig. \ref{figSED}. The correction factor for each filter is computed following \citet{Roming09}.\\

In infrared and optical light, the SED of \onees\ is dominated by the host galaxy, a giant elliptical well-studied in the past \citep{Scarpa00, Urry00, Nilsson03}. For the purpose of subtracting the host-galaxy contamination, we made use of the recent results from \citet{Nilsson07}, who estimate a contaminating flux of $1.01 \pm 0.06$ mJy in the R band, for an aperture radius of 5". This contribution is then translated into a correction in the V, B and U filters following \citet{Hyvonen07}. An uncertainty in the host-galaxy color of $0.1$ is considered, and included in the final error bars of the \uvot\ spectral points. For the remaining \uvot\ filters we assume that the host-galaxy contribution is negligible compared to the AGN.\\
 
\subsection{\flwo}
As part of a long-term optical program of monitoring of VHE blazars, \onees\ is regularly observed by the automatic \flwo\ telescope, located near the \veritas\ site.\footnote{\url{http://www.sao.arizona.edu/FLWO/48/48.html}}  The instrument, a reflector of 1.2-m diameter, can perform measurements using the standard Cousins and SDSS filters. The analysis uses an aperture radius of 10" to estimate the source magnitude.  There are no observations performed simultaneously with \veritas: the observation closest to the \veritas\ detection of \onees\ was performed on May 18, 2013, eleven (ten) days after the last detection by \veritas\ (\swift), using B, r', and i' filters. The data are dereddened using $E_{B-V}=0.037$, as for \uvot, and the host-galaxy contribution is subtracted following \citet{Nilsson07} and translated into the relevant filters using \citet{Hyvonen07} (for the B filter) or \citet{Fukugita95} (for the remaining SDSS filters). Similar to \uvot, the uncertainty in the dereddening and the host-galaxy color is taken into account and included in the error bars. The three  \flwo\ measurements, corrected for both absorption and host-galaxy contamination, are plotted in Fig.~\ref{figSED}.\\

\section{SED modeling}

 \begin{table}
   \centering
   \caption{Parameters used for the SSC modeling of \onees.   	\label{table1}}
   		\begin{tabular}{l c }
		\hline
		\hline
		Parameter & Value \\
		\hline
		$\delta$ & \textit{30} \\
		\hline
		$\gamma_{min}$ [$10^3$]& \textit{1-5}  \\
		$\gamma_{break}$ [$10^5$]& $2.2$-$3.2$\\
		$\gamma_{max}$ [$10^6$]& $5.5$-$7.0$\\
		$\alpha_{e,1}$ &\textit{2.2}\\
		$\alpha_{e,2}$ &\textit{3.2}\\
		$K'_e$ [$10^{-11}\ \textrm{cm}^{\textrm{-3}}\textrm{]}$ & $3.5$-$29.3$  \\
		$^\star u_e$ [$10^{-5}\ \textrm{erg cm}^{\textrm{-3}}\textrm{]}$ & $1.6$-$12.8$ \\
		$^\star u_{ph}$ [$10^{-9}\ \textrm{erg cm}^{\textrm{-3}}\textrm{]}$ & $8.5$-$41.5$ \\
		\hline
		$R_{src}$ [$10^{17}$ cm] & $4.3$-$7.4$\\
		$^\star \tau_{var} [days]$ & $5.8$-$10.0$ \\
		\hline
		$B$ [mG] & $0.3$-$0.6$\\
		$^\star u_B$ [$10^{-9}\ \textrm{erg cm}^{\textrm{-3}}\textrm{]}$ & $3.6$-$16.2$\\
		\hline
		$^\star u_e/u_B$ [$10^3$] & $1.6$-$13.8$  \\
		$^\star u_{ph}/u_B$  & $1.2$-$2.8$ \\ 
		$^\star L_{jet}$ [$10^{44} \textrm{erg s}^{\textrm{-1}}\textrm{]}$& $1.4$-$5.3$\\
		\hline
		\hline		
		\end{tabular}
		\vspace{0.2cm}
				\tablecomments{Italicized values are fixed in the fitting algorithm. Derived quantities are marked with a star. Two different values of $\gamma_{min}$ are studied, and do not affect the other free parameters. The values of $\gamma_{max}$ are simply equal to $22\times\gamma_{break}$. \vspace{0.5cm}}	
		\end{table}

\label{section5}
\begin{figure*}[hbtp]
\begin{center}
\includegraphics[width=450pt]{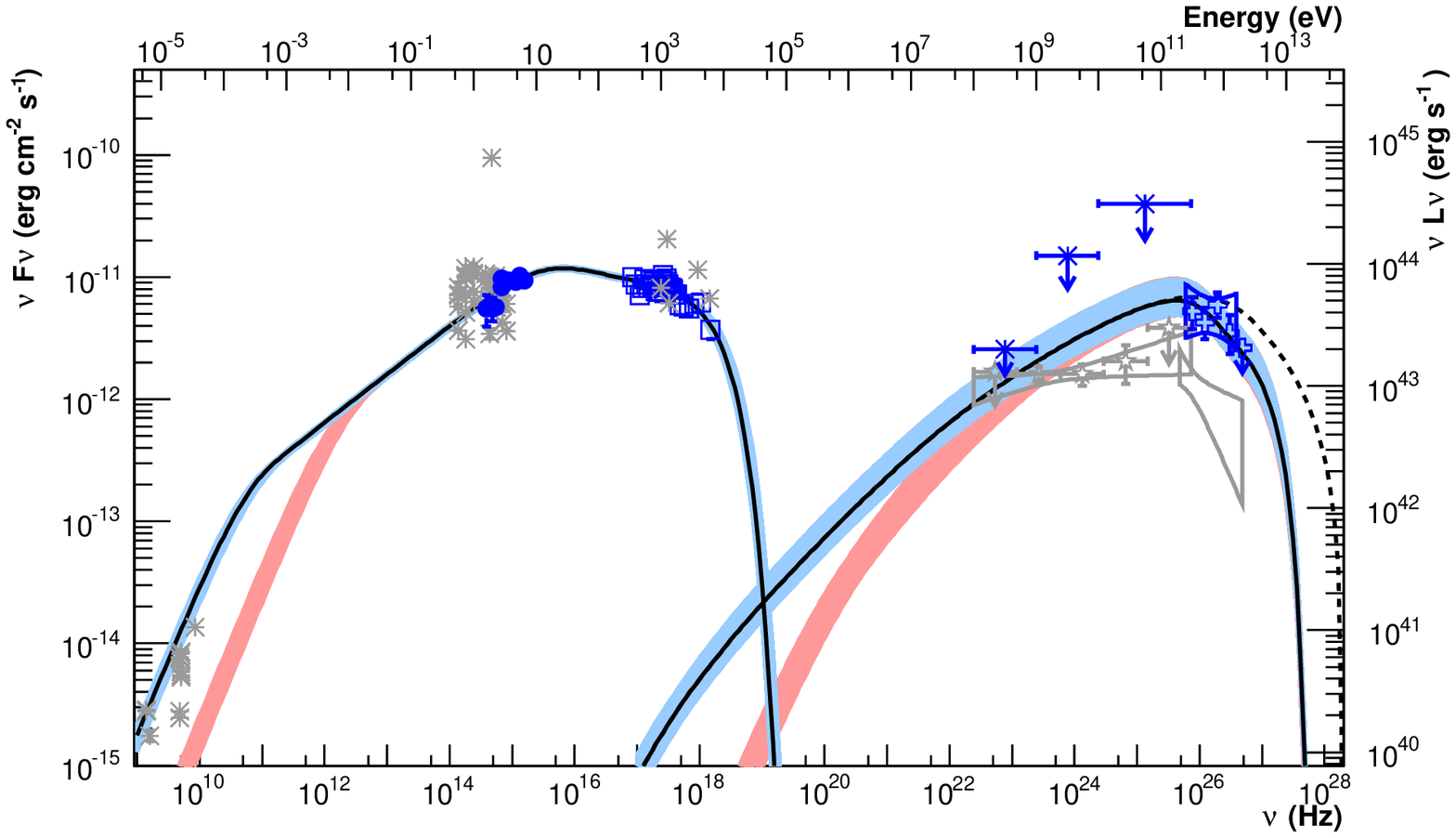}
\caption{Quasi-simultaneous SED of \onees\ (in blue; from low to high energies: \flwo, \swiftuvot, \swiftxrt, \fermilat\ and \veritas). Archival data (from the NED, \fermilat, and \magic) are included, and plotted in gray. The light-blue and pink curves represent the two sets of SSC models computed assuming $\gamma_{min}=10^3$ and $5\times10^3$ (see Table \ref{table1} and Section \ref{section5}). The black model represents the solution which minimizes the $\chi^2$ value. The black dotted line represents the same model without absorption on the EBL, estimated using the template by \citet{Franceschini08}. The luminosity ($y$-axis at the right) is computed assuming $z=0.055$ and the  cosmological values (flat universe, with $H_0=67.3$ km s$^{-1}$ Mpc$^{-1}$ and $\Omega_m=0.315$) estimated using the measurements made by the \textit{Planck} satellite  \citep{Planck13}.}
\label{figSED}
\end{center}
\end{figure*}

The SED of \onees\ is shown in Fig. \ref{figSED}. Simultaneous observations (with \swiftxrt, \swiftuvot, and \veritas) were carried out only during MJD 56419. Given that none of the instruments involved in the campaign detected any significant variability, we make the assumption that the flux was constant during the observations, and we thus consider that the average spectral measurements from the different instruments are representative of the blazar emission. It is possible that the emission was variable, but at a level below instrumental sensitivities. Nonsimultaneous \flwo\ measurements also cover the infrared-to-optical part of the SED. In this case as well, given the absence of variability in the B filter between \swiftuvot\ and \flwo\ observations, we make the assumption that the \flwo\ measurements can be considered as representative of the emission from \onees\ during the \veritas\ observations. \fermilat\ upper limits calculated from observations performed during the \veritas\ campaign (but not strictly simultaneously) are also included, as well as the average spectrum from the long-term analysis. The bow-tie from the \magic\ detection and the archival MWL data from the NED (NASA/IPAC Extragalactic Database)\footnote{\url{http://ned.ipac.caltech.edu}} are also included.\\

In the context of the synchrotron-self-Compton (SSC) model, the two components of the blazar SED are associated respectively with synchrotron emission from leptons (e$^\pm$) and inverse-Compton scattering of the particles off synchrotron radiation from the same lepton population. The emitting region is a spherical blob of plasma (characterized by its radius $R$) in the relativistic jet, moving towards the observer with Doppler factor $\delta$, and filled with a tangled, homogeneous magnetic field $B$. The particle population is parametrized by a broken power-law function, and it carries six free parameters: 
the minimum, maximum, and break Lorentz factors of the particles ($\gamma_{min}$, $\gamma_{max}$, $\gamma_{break}$), the two indices ($\alpha_1$ and $\alpha_2$), and the normalization factor $K$. The minimum Lorentz factor of the leptons can be fixed at a reasonably low value without affecting the modeling. The remaining eight free parameters can be constrained by observations, as discussed, for example, in \citet{Bednarek97, Tavecchio98} and \citet{Cerruti13}.\\

 Given the lack of a simultaneous \fermilat\ detection, it is impossible to constrain the position and the luminosity of the inverse-Compton peak during the high state, and a unique solution for the SSC model cannot be provided.  However, the synchrotron component is very well sampled, and can provide some constraints  on the energy distribution of particles in the emitting region.
At low energies, the subtraction of the host-galaxy contamination reveals the AGN nonthermal continuum, which can be described by a power law from infrared to UV, with no sign of break. A fit of the \flwo\ and \uvot\ data results in an index  $n_1=1.64\pm0.09$. In the case of synchrotron radiation, this index reflects directly the index of the underlying e$^\pm$ population $\alpha_1=2n_1-1$, which is thus equal to $2.28 \pm 0.18$. Similarly, the index of the X-ray power law below the X-ray break corresponds to $\alpha_2=3.0\pm0.2$. In the following, $\alpha_1$ has been fixed to $2.2$ and $\alpha_2$ to $3.2$, i.e. a spectral break of $1.0$ has been assumed.\\

In blazar physics, an additional observational constraint comes from the variability timescale $\tau_{var}$ and the causality argument: the emitting region size $R$ has to be smaller than $c \tau_{var} \delta/(1+z)$. However, for the case of \onees, there is no evidence of variability on short timescales in any of the light curves that can provide an estimation of $\tau_{var}$ and thus constrain the emitting region size. \\

As discussed above, the value of $\gamma_{min}$ does not affect the modeling and it is therefore held fixed. However, values lower than $10^3$ would overestimate archival radio measurements (see Figure \ref{figSED}), even though, given their nonsimultaneity, these data should not be considered as a strong constraint. An additional constraint on $\gamma_{min}$ is provided by the \fermilat\ nondetection, and will be discussed in the next section. In the following we study two different cases, for $\gamma_{min}=10^3$ and $5\times10^3$. The value of $\gamma_{max}$ is constrained by the break observed in the X-ray spectrum of \onees. In the following, we express it as a function of $\gamma_{break}$: $\gamma_{max}=22\ \gamma_{break}$. The numerical factor corresponds to the ratio between the X-ray break and the synchrotron peak, which is estimated by extrapolating the UV and X-ray spectra.  \\

Once the indices and maximum energy of the particle population have been fixed, the number of free parameters in the SSC model is thus five ($\delta$, $B$, $R$, $\gamma_{break}$, and $K$), and, with only four observables (the frequency and flux of the synchrotron peak, the VHE spectral index and the VHE flux at 620 GeV: $\nu_{syn-peak}$; $\nu F_{\nu;syn-peak}$; $\Gamma_{VERITAS}$ and $\nu F_{\nu;VERITAS}$), a unique solution cannot be provided. However, assuming a reasonable value of the Doppler factor, it is possible to study the parameter space of the remaining free parameters. The best-fit solution is computed using the numerical algorithm described in \citet{Cerruti13}. The parameter space is systematically sampled, producing a set of SSC models and computing for each of them the expected observable values. A fit is then performed in order to express each observable as a function of the free parameters, defining a set of equations which is solved for the particular set of observables of \onees. The uncertainty in the observables is taken into account by iteratively solving the system of equations, producing a family of solutions and determining the allowed range for each free parameter.\\

For the particular case of \onees, the Doppler factor has been fixed to $30$ \citep[a value in line with standard SSC modeling of VHE blazars, see e.g.][]{Tavecchio10, Zhang14}, while the other free parameters have been studied in the range: $B \in [0.1, 2]$ mG; $R \in [6\times10^{16}, 10^{18}]$ cm; $K' \in [5\times10^{-12}, 5\times10^{-10}]$ cm$^{-3}$; $\gamma_{break} \in [10^5,10^6]$. $K'$ is defined as the particle density at $\gamma_{break}$: $K'=K \gamma_{break}^{-\alpha_1} $.
The solutions of the SSC model are computed iterating on the four observables: $\nu_{syn-peak} \in [10^{15.81},10^{15.85}]$;  $\nu F_{\nu;syn-peak} \in [10^{-10.94},10^{-10.92}]$; $\nu F_{\nu;VERITAS} \in [10^{-11.46},10^{-11.18}]$ and $\Gamma_{VERITAS} \in [1.78,2.52]$. For the two \veritas\ observables, the systematic uncertainty has been taken into account and summed in quadrature with the statistical error. The system of equations is completed by an inequality relating the variability timescale to $R$ and $\delta$. We considered $\tau_{var}=10$ days, which corresponds roughly to the interval between the last \veritas\ detection on MJD 56419 and the observation during MJD 56430, which indicated that the VHE flare may have ended.  All the SSC models that correctly describe the \onees\ SED are then recomputed, and plotted in Fig. \ref{figSED}. The parameter values are listed in Table \ref{table1}, together with derived quantities such as the energy densities, the luminosity of the emitting region and the minimum variability timescale. For each solution, the $\chi^2$ with respect to the observational data is calculated in order to find the solution that minimizes the $\chi^2$. \\

\section{Discussion}
\label{section6}
The \veritas\ detection of VHE $\gamma$-ray emission from \onees\ triggered MWL observations that allowed, for the first time, the study of the SED during a high $\gamma$-ray flux state, a factor of five higher than the \magic\ detection.\\
 Even though the lack of simultaneous detection by \fermilat\ does not allow the parameter space to be fully constrained, the available data permit some important conclusions to be drawn regarding the properties of the particle population and the acceleration/cooling mechanisms.\\  

In the presence of synchrotron radiation, a break in the stationary particle population is expected, and it is characterized by $\Delta \alpha=1$ \citep[see e.g.][]{Susumu}. Deviations from $\Delta \alpha=1$ can, however, occur if the emitting region is inhomogeneous, or if the emission from escaped particles is taken into account \citep[see e.g.][]{Sokolov04, Chen14}. Interestingly, the measurements of the synchrotron component of \onees\ are consistent with a simple break of 1.0, indicating that a single particle population, injected with a power-law distribution with index equal to $2.28\pm0.18$, could explain the blazar emission.\\

 The break energy corresponds to the equality of the escape (or adiabatic) timescale and the synchrotron and inverse-Compton cooling timescales. The former is energy independent (if the escape is advective, and not diffusive), and equal to $R /\beta c$, while the cooling timescales are both proportional to $1/\gamma$.
 Following \citet[][equation 30]{Tavecchio98}, it is possible to relate $\gamma_{break}$ to the model parameters and find the only unknown variable, $\beta \in [1/225, 1/70]$. This means that the stationary particle population is consistent with an injected power-law particle population cooled by synchrotron and inverse-Compton emission only if the advective escape timescale is $\sim 100 R/c$. This value is low compared to the ones used by \citet{Tavecchio98} ($\beta \in [0.33, 1]$). For example, assuming $\beta=1$, the break energy would have been at $\gamma_{break} \sim 10^{7}$, two orders of magnitude higher than the constraint provided by quasi-simultaneous observations. Alternatively, the hypothesis of a power-law population injected and cooled by synchrotron and inverse-Compton processes could be too simplistic, and additional injection/cooling/escape terms could contribute to the final stationary particle population.\\
 
It is important to recall that these considerations are valid only for a given value of $\delta=30$. Lower values of $\delta$ would imply higher values of $B$ and then higher values of $\beta$. For example, assuming $\delta^\star=15$, the overall normalization is reduced by a factor of $2^4=16$ and can be compensated by assuming a magnetic field four times stronger, $B^\star \sim 1.6$ mG. In order to maintain the same $\nu_{syn-peak}$ (which is $\propto \delta B \gamma_{break}^2$), $\gamma_{break}$ has to be reduced by a factor of $\sqrt{2}$. These transformations would affect the inverse-Compton peak as well, and it is not guaranteed that these parameters could reproduce the VHE spectral index. Regardless, using these new values, $\beta^\star \propto \gamma_{break}^\star B^{\star 2} \simeq 11 \beta$, one order of magnitude higher than that computed for $\delta=30$.\\ 

It is possible to extract information on the acceleration mechanism from the value of the index of the particle population.  Acceleration by diffusive shocks naturally produces power-law particle populations, with an index close to 2.0. For ultrarelativistic shocks, several authors \citep[see][and references therein]{Achterberg01} have shown that $\alpha_1$ is expected to be around 2.2. The quasi-simultaneous multiwavelength (from infrared to UV) observations of \onees\ show that $\alpha_1$ has to be between $2.10$ and $2.46$, fully consistent with ultrarelativistic shock acceleration.\\

The maximum particle energy is another important model parameter related to the acceleration.  The break observed in X-rays at $\simeq 1.2$ keV can indeed be related to the high-energy cutoff of the particle population ($\gamma_{max}$), and can thus be used to constrain the maximum energy of the leptons. In the framework of the standard diffusive shock acceleration, the maximum particle energy can be computed assuming that the acceleration timescale equals that for the radiative losses: particles with energy higher than $\gamma_{max}$ lose energy faster than they can be accelerated. Assuming $\tau_{acc} = \eta\ mc\gamma/eB=(5.7\times10^5 {\rm s})\ \eta\gamma/B$, where $\eta$ represents a parameter characteristic of the shock acceleration, it is easy to estimate $\gamma_{max} = 3.7\times10^9\ (\eta B_{mG})^{-0.5}$, which is consistent with the constrained values of $\gamma_{max}$ only for $\eta$ of the order of $10^5$. Such a high value, which implies rather inefficient acceleration, is typical of high-frequency-peaked BL Lac objects \citep[see e.g.][]{Susumu}. In this case, solutions computed for lower values of $\delta$ cannot significantly lower the value of $\eta$. It is easy to show that, using the same values estimated in the discussion on $\gamma_{break}$, $\eta$ would be reduced only by a factor of two.\\ 
   
   The minimum particle energy  $\gamma_{min}$ also carries important information on the physics of the emitting region. Both archival radio measurements and the nondetection by \fermilat\  suggest that $\gamma_{min}$ has to be at least of the order of $10^3$. The study of the models for two different values of $\gamma_{min}$ (equal to $1$ and $5\times10^3$) shows that solutions with lower values of $\gamma_{min}$ have higher emission in the 0.1-1 GeV energy band. A $\gamma_{min}$ lower than $10^3$ would start conflicting with the \fermilat\ nondetection (under the assumption that the \veritas\ flux is representative of the average flux throughout the whole period used to compute the \fermilat\ upper limits). Such a high value of $\gamma_{min}$ is not a surprise for HBLs \citep[see][]{Aliu0229}. It can be explained, for example, by assuming that the particles are already injected with a truncated-power-law distribution, and they do not have time to cool down completely, or that the particles are somehow reaccelerated inside the blazar region \citep{Kata06}. \\ 
   
 It is also interesting to note that the one-zone SSC modeling suggests that the blazar GeV spectrum during the high-flux state was harder compared to the average spectrum measured by \fermilat. A harder-when-brighter correlation in the \fermilat\ band is not common for HBLs \citep[see for example][]{Mrk421, Mrk501}, and is more similar, for example, to what is observed during the $\gamma$-ray flaring activity of BL Lacertae \citep{Arlen13}.\\
   
 One of the open questions in blazar physics is related to the energy budget of the emitting region. Is it close to equipartition between particle, photon, and magnetic energy densities ($u_e$, $u_{ph}$ and $u_B$)? The solutions estimated here are all characterized by a very high value of $u_e/u_B$, which is of the order of $10^{3-4}$. On the other hand, the photon energy density (synchrotron emission by primary leptons) is of the same order as the magnetic one. Equipartition is sometimes useful to constrain the parameter space of blazar models, especially when dealing with external photon fields, which increase the number of free parameters. Recently \citet{equipartition1} presented a general approach to compute blazar models close to equipartition, showing that, within this framework, it is possible to describe the SED of blazars, and in particular FSRQs, correctly fitting the spectral break observed by \fermilat\ in several low-frequency-peaked blazars \citep{equipartition2}. On the other hand, it is not clear if a general equipartition approach still holds for HBLs, and indeed several modeling attempts (within a standard one-zone SSC scenario) on the most well studied HBLs result in equipartition factors far from unity \citep[see][]{Mrk421, Mrk501, pksmwl}.\\
 
 It is important to emphasize that the SED of \onees\ represents a snapshot of a high-flux state, and that the ratio between the flaring and the archival fluxes is higher in the inverse-Compton component than in the synchrotron one. It is thus possible that the emitting region is only temporarily far from equipartition, and that the low state is indeed characterized by $u_e/u_B$ closer to one. To test this hypothesis, the constraining algorithm has been rerun assuming $\nu F_{\nu;VERITAS}$ lower by a factor of five (i.e. consistent with the \magic\ detection). As expected, the equipartition factor is lowered by one order of magnitude, lying between $50$ and $1000$. Again, solutions computed for lower values of $\delta$, implying a higher value of $B$, would result in $u_e/u_B$ much closer to equipartition.
In addition, since the particle index $\alpha_1$ is softer than $2.0$, as constrained by optical/UV observations, the value of $u_e$ is strongly dependent on $\gamma_{min}$. Given that we cannot really constrain the value of $\gamma_{min}$, it is possible that the equipartition factor is indeed closer to unity.\\
 
\section{Conclusions}
\label{section7}

In this paper we presented the results of a multiwavelength campaign carried out on the BL Lac object \onees\ during May 2013, triggered by a VHE high-flux state detected by \veritas.
This represents the first detection of a blazar flare with \veritas\ during bright moonlight observations.\\
 
 Within this campaign, no significant variability is detected at VHE, nor at lower energies (X-rays and optical). The \veritas\ light curve is consistent with a constant flux ($6.3\%$ Crab above 250 GeV, roughly five times the archival \magic\ detection) between May 01, 2013 and May 07, 2013; additional observations during May 18, 2013 indicate (at a $2.4\sigma$ level) that the high-flux state may have ended at some point after the last \veritas\ detection on May 07.\\ 
 
  The quasi-simultaneous SED is fitted  by a standard one-zone SSC model. Even though the nondetection by \fermilat\ did not enable a full study of the parameter space, the measurements are fully consistent with particle acceleration by relativistic diffusive shocks and simple synchrotron and inverse Compton cooling, resulting in a power law with injection index around 2.2 and a spectral break of 1.0.\\
 
In recent years the number of VHE blazars has significantly increased thanks to the current generation of Cherenkov telescope arrays such as \veritas , \magic\ and \hess. The broadband emission and the rapid variability of blazars require prompt simultaneous multiwavelength campaigns to fully understand the physics of their emitting region. The new \veritas\ observing strategy under bright moonlight will be particularly useful for blazar science, significantly increasing blazar monitoring capabilities at VHE.\\

Combining data sets taken under various observing conditions is not a new aspect of VHE gamma-ray data analysis. In the case of \veritas, lookup tables for the analysis are generated from simulations which already assume a wide range of different zenith angles, azimuth angles (and hence geomagnetic field strengths), sky brightness and local atmospheric conditions. The introduction of a new observing mode for \veritas\ is therefore not expected to adversely affect the scientific interpretation of the results. Exploring new modes is certainly worthwhile, as the results of this paper demonstrate. Bright moonlight observing time is particularly useful for variable VHE sources, and can also be used to obtain deeper exposures on steady sources, especially for targets with hard VHE spectra which can be detected despite an increased energy threshold. The merging of data taken under multiple observing modes and detector configurations will be an important analysis challenge for future large arrays, such as the Cherenkov Telescope Array (\textit{CTA}), particularly with regard to deep VHE exposures and extended surveys.

\footnotesize{
\acknowledgments{The authors wish to thank Kari Nilsson for useful discussions about the host-galaxy contribution, as well as the anonymous referee for his/her comments which improved the present work. This research is supported by grants from the U.S. Department of Energy Office of Science, the U.S. National Science Foundation and the Smithsonian Institution, by NSERC in Canada, by Science Foundation Ireland (SFI 10/RFP/AST2748) and by STFC in the U.K. We acknowledge the excellent work of the technical support staff at the Fred Lawrence Whipple Observatory and at the collaborating institutions in the construction and operation of the instrument. The VERITAS Collaboration is grateful to Trevor Weekes for his seminal contributions and leadership in the field of VHE gamma-ray astrophysics, which made this study possible.\\}
}

\end{document}